\documentclass[aip,amsfonts,amsmath,amssymb,reprint]{revtex4-2}
\usepackage{graphicx}
\usepackage{dcolumn}
\usepackage{bm}

\usepackage{xcolor}
\usepackage[english]{babel}
\usepackage[utf8]{inputenc}
\usepackage[T1]{fontenc}
\usepackage{mathptmx}
\usepackage{tabularx,booktabs}
\usepackage{soul}
\usepackage{color}

\setstcolor{red}

\newcommand{\Pm}[1]{$P(m;t,N)$}

\begin{document}

\preprint{----}

\title{A hybrid percolation transition at a finite transition point in scale-free networks}

\author{K. Choi}
\author{Wonjun Choi}%
\author{B. Kahng}%
\email{bkahng@snu.ac.kr}
\affiliation{ 
CCSS, CTP, Department of Physics and Astronomy, Seoul National University, Seoul 08826, Korea
}%

\date{\today}

\begin{abstract}
Percolation transition (PT) means the formation of a macroscopic-scale large cluster,  which exhibits a continuous transition. However, when the growth of large clusters is globally suppressed, the type of PT is changed to a discontinuous transition for random networks. A question arises as to whether the type of PT is also changed for scale-free (SF) network, because the existence of hubs incites the formation of a giant cluster. Here, we apply a global suppression rule to the static model for SF networks, and investigate properties of the PT. We find that even for SF networks with the degree exponent $2 < \lambda  <3$, a hybrid PT occurs at a finite transition point $t_c$, which we can control by the suppression strength. The order parameter jumps at $t_c^-$ and exhibits a critical behavior at $t_c^+$. 
\end{abstract}

\maketitle

\begin{quotation} 
Percolation is a useful tool to understand diverse phenomena on and of complex networks such as resilience, epidemic spreading, and community formation. Unfortunately, however, the percolation threshold $t_c$ of random scale-free networks with degree exponent $2 < \lambda < 3$ is zero in the limit $N\to \infty$, where $N$ is the system size. Here, we show that a transition point $t_c$ can be delayed by controlling the suppression strength against the growth of large clusters. For this case, a hybrid transition occurs with more interesting properties: the order parameter jumps and then exhibits critical behaviors at $t_c$.
\end{quotation}

\section{Introduction}
Percolation is a geometric phase transition from a non-percolating state to a percolating state~\cite{stauffer}. For instance, let us assume a square lattice in which each bond is occupied with probability $p$. When $p$ is smaller (larger) than a characteristic value $p_c$, a spanning cluster is not formed (is formed) between two opposite sides of the lattice. A percolation transition (PT) occurs at the transition point $p_c$. This behavior is not limited to systems in Euclidean space but extends to complex networks. At each time step, a link is added between two nodes randomly selected from $N$ initially isolated nodes. When the number of links added exceeds a characteristic value $L_c$, a giant cluster of size $O(N)$ is formed. This process is described by the Erd\H{o}s--R\'enyi (ER) model~\cite{Erd}. The notion of percolation has long served as a basic paradigm for understanding network resilience, community formation, the spread of disease in a population, and so on
~\cite{resilience2,cohen00,redner,mendes_percolation,lee04}.

A PT is generally considered to be one of the most robust continuous transitions; however, PTs in complex systems often exhibit various types such as continuous, discontinuous, hybrid, and infinite-order phase transitions~~\cite{DSouza2015,ziff_rev,lee_rev,diverse,callaway}. Thus, extension of the original PT model has received considerable attention in recent years. There was a trial to break the robustness of continuous PTs: A random network model in the so-called Achlioptas process was proposed; this model is referred to as an explosive percolation (EP) model~\cite{Achlioptas2009}. In the Achlioptas process, two pairs of nodes that are not yet connected are chosen randomly, and the pair that produces a smaller cluster is taken for connections. Thus, the Achlioptas process suppresses the growth of large clusters; consequently, the percolation threshold is delayed, and medium-size clusters become abundant in the system before the threshold. As a result, when the percolation threshold is crossed, the size of the largest cluster is dramatically increased by the coalescence of these medium-size clusters in a short time scale~\cite{DSouza2015,cho_supp}. Thus, the PT in the EP model was regarded as a discontinuous transition when it was first proposed; however, the suppression is too local, and the PT in the EP model was found to be continuous in the thermodynamic limit~\cite{mendes}. The robustness of the continuous PT was reinforced rather than weakened. 

Extensive studies of the Achlioptas process revealed that a global suppression rule is necessary to generate a discontinuous PT~\cite{riordan}. Subsequently, a solvable model in two dimensions was introduced~\cite{Cho2013}. At each time step, this model checks whether a bond to be added generates a spanning cluster. If a spanning cluster is formed, the candidate bond is not allowed for occupation. Checking for a spanning cluster naturally requires global information and thus meets the criterion for a discontinuous PT. However, this model does not exhibit any critical behavior because the PT is a generic  discontinuous transition.

A hybrid PT (HPT) is another type of PT in which the order parameter jumps but criticality appears at a transition point. The so-called restricted ER ($r$-ER) model  proposed in Ref.~\cite{Panagiotou2011} exhibits such features. We modified the original model slightly~\cite{Cho2016}. In this model, clusters are partitioned into two sets; set $A$ contains the smallest clusters, and set $B$ contains the remaining large clusters. The number of nodes in set $A$ is limited to $gN$, where $g$ represents a fraction and is controllable, and $N$ is the number of nodes in the system. At each time step, one node is selected randomly from all the nodes, and another node is selected from the restricted set $A$. These nodes are connected, and their clusters are merged. As the sizes of these clusters change, the elements of each partition may be updated. Global information is acquired during this updating process. This $r$-ER model exhibits an HPT in which the order parameter jumps from zero to a finite value at a transition point, and the size distribution of the finite clusters exhibits power-law decay, indicating underlying critical behavior~\cite{Cho2016}. We recently found that the underlying mechanism is tug-of-war-type criticality~\cite{Park2019}. Hence, the PT transition is hybrid. 

A question naturally arises as to whether an HPT occurs in a similar restricted scale-free ($r$-SF) network model. Here, a scale-free network is a network in which the degree distribution follows a power law expressed as $P_d(k)\sim k^{-\gamma}$. This question may be interesting, because two competing effects are present in $r$-SF networks. One is induced by the presence of hubs, which encourages the formation of a giant cluster. The other is induced by the restriction rule, which discourages the formation of large clusters. Owing to the resulting competition, the evolution of scale-free networks under the half-restricted rule would be nontrivial. Here, we investigate a PT emerging from the $r$-SF rule. 

This paper is organized as follows. In Sec.~\ref{sec:RSM}, we propose the $r$-SF model, which is a type of cluster-merging percolation model. In Sec.~\ref{sec:DT}, we perform numerical simulations for the $r$-SF model and analyze the numerical results. It is shown that the order parameter of the $r$-SF model exhibits a jump as $t$ approaches $t_c$. 
In Sec.~\ref{sec:CH}, we obtain the values of the critical exponents related to the critical phenomena in the HPTs that occur on complex networks, including scale-free networks. Finally, in Sec.~\ref{sec:summary}, we summarize our results and discuss the implication of this study.

\section{Restricted scale-free network model}\label{sec:RSM}

Motivated by the $r$-ER model, we propose an $r$-SF model, which is a modification of the static model~\cite{static_model}. The original static model~\cite{static_model,lee04} is defined as follows.  
Consider a system of $N$ nodes indexed as $i=1,\cdots, N$. Each node $i$ is assigned a weight $i^{-\mu}$, where $\mu$ is a real number in the range $\mu \in [0,1]$.
At each time step, two nodes $i$ and $j$ ($i\ne j$) are chosen with probabilities $p_i$ and $p_j$, respectively, which are given by
\begin{align}
	p_{i} = \frac{i^{-\mu}}{\zeta_{N}(\mu)}, \label{eq:prob}
\end{align}
where $\zeta_{N}(\mu) =\sum_{i=1}^{N}i^{-\mu}$, and $p_{j} = {j^{-\mu}}/{\zeta_{N}(\mu)}$.
The nodes are connected unless they are already connected. The resulting network is a scale-free network, and its degree distribution follows the power law $P_d(k)\sim k^{-\lambda}$, where $\lambda=1+1/\mu$, because $\mu \in (0,1]$, $\lambda \ge 2$.  

\begin{figure}[!t]
\centering
\includegraphics[width=1.0\linewidth]{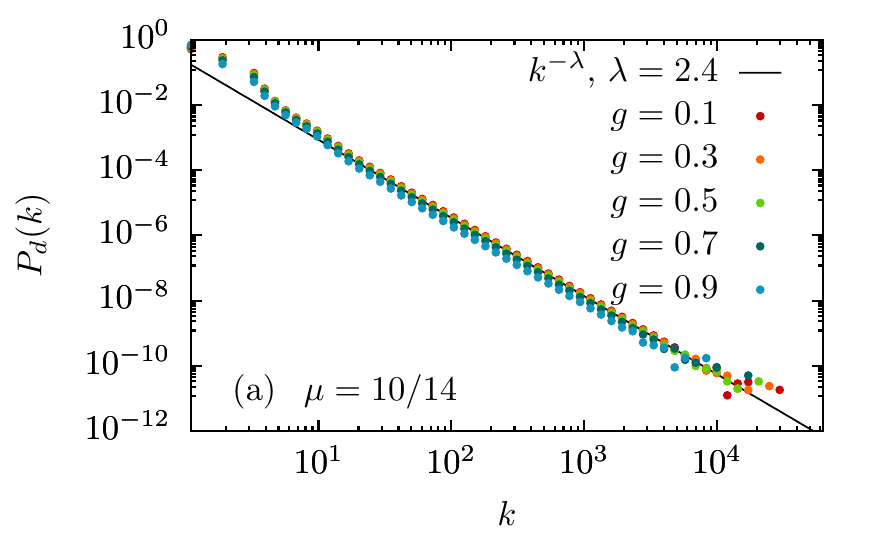}
\includegraphics[width=1.0\linewidth]{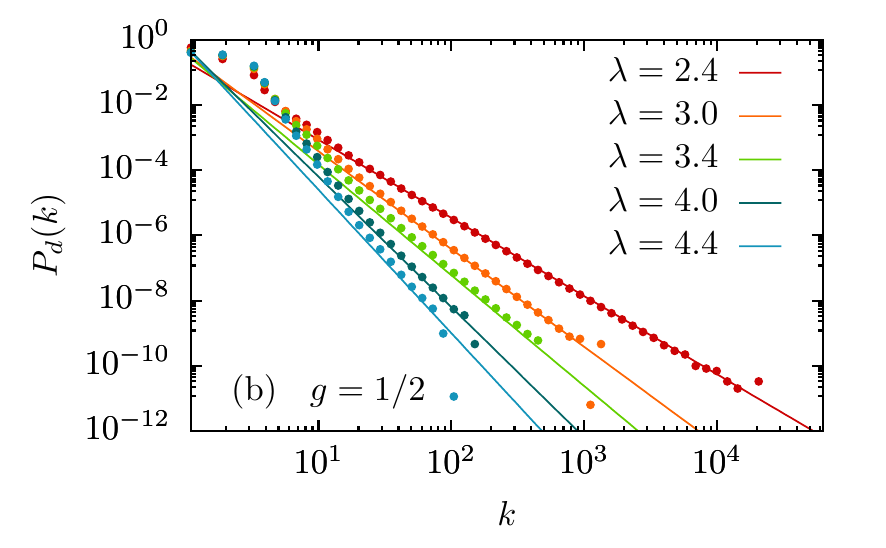}
\caption{
(a) Plot of the degree distribution $P_{d}(k)$ for various $g$ and a fixed $\mu=5/7$ at $t_c(g)$. As $g$ is decreased, $t_c$ is more delayed as shown in Fig.~2(b). Thus, more links are added to the system up to $t_c$ and  the degree distribution is more extended. (b) Plot of the degree distribution for various $\mu$ but a fixed $g=1/2$ at $t_c$. The guide lines for each case have the slope $\lambda = 1+1/\mu$ theoretically predicted.
}
\label{fig:dd}
\end{figure}

For the $r$-SF network model, we classify clusters into two sets, $A$ and $B$, in a way similar to that used in the $r$-ER model. Set $A$ contains approximately $gN$ nodes belonging to the smallest clusters. The remaining approximately $(1-g)N$ nodes belong to set $B$. More rigorously, 
clusters are ranked in ascending order of size. The $i$-th cluster and its size are denoted as $c_{i}$ and $s(c_{i})$, respectively. Clusters with the same size are randomly ordered. We define a cluster $c_{k}$ that satisfies $\sum_{i=1}^{k-1}s(c_{i}) < \left\lfloor gN \right\rfloor \leq \sum_{i=1}^{k}s(c_{i})$; then we determine set $A$ as $ \left\{c_{1},\ldots, c_{k}\right\}$, where $g\in (0,1]$ is a real number.

Initially the system is composed of $N$ isolated nodes. We begin by selecting two nodes; one node $i$ is selected from among all the nodes with the probability $p_{i}$ given in Eq.~\eqref{eq:prob}.  
The other node is selected from set $A$ with probability $p^{\prime}_{j}$, which is given as 
\begin{align}
	p^{\prime}_{j} = \frac{r_{j} j^{-\mu}}{\sum_{k=1}^{N}r_{k} k^{-\mu}},
\label{eq:fitness}	
\end{align}
where $r_{i}=1$ if node $i$ is in set $A$; otherwise, $r_{i}=0$. The nodes are connected unless they are already connected. We repeat this process until the system has $L$ links. The model parameter $g$ determines the cluster size $s(c_k)$ at the boundary between sets $A$ and $B$. When $g=1$, the partition does not exist, and the $r$-SF model reduces to the original static model. However, when $g\to 0$, most clusters belong to set $B$, and the growth of most clusters is suppressed. Thus, $1-g$ represents the suppression strength effectively. 
For the $r$-SF model, even though the system is partitioned into two parts, the degree distribution keeps the same form independent of the parameter $g$, as shown in Fig.~\ref{fig:dd}. However, it is more extended as $t_c$ is more delayed for smaller $g$. So the maximum degree denoted as $k_{\rm max}$ becomes large. This is because as time is related to the mean degree as $t= \int_0^{k {\rm max}} k P_d(k) dk/(2N)$ in the subcritical regime.

\section{Discontinuity of the order parameter}\label{sec:DT}

\begin{figure}[!tbp]
\centering
\includegraphics[width=1.0\linewidth]{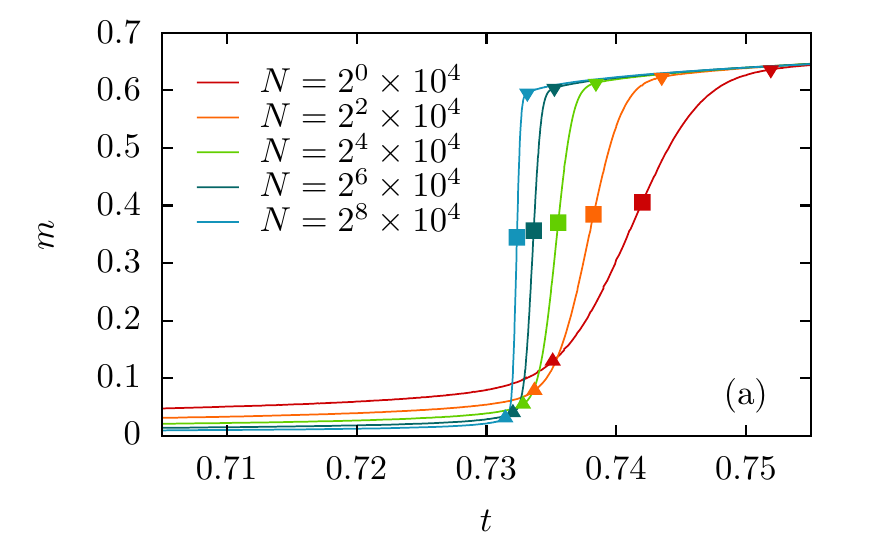}
\includegraphics[width=1.0\linewidth]{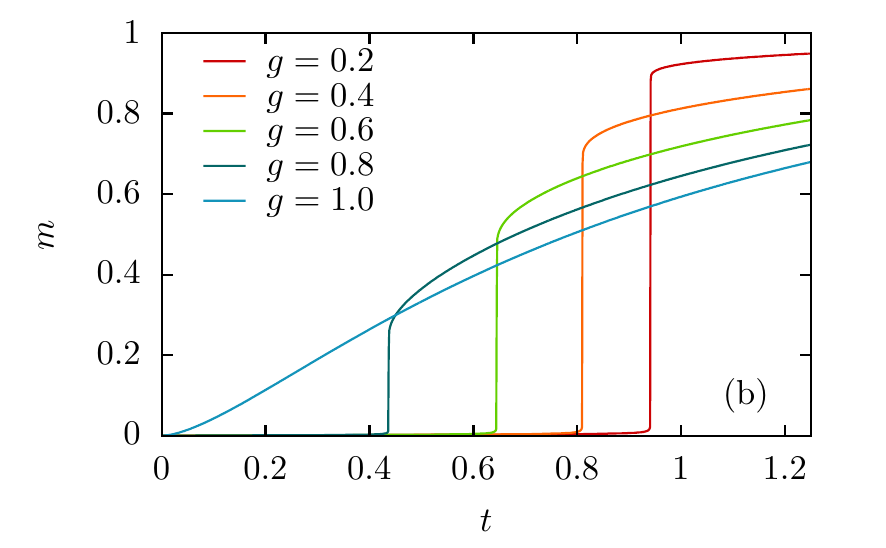}
\caption{
(a) Plot of the order parameter $m(t;N)$ as a function of time $t\equiv L/N$ for different system sizes. Several characteristic times are indicated: the upper transition point $t_c^+(N)$ ($\blacktriangledown$), the lower transition point $t_{c}^{-}(N)$ ($\blacktriangle$), and the point at which the slope of $m(t)$ becomes maximum, $t^{*}(N)$ ($\blacksquare$). They are estimated as $t_c^+(N)\approx 0.731924$, $t_{c}^{-}(N)\approx 0.731023$, and $t^*(N)\approx 0.731468$ for $N=2^{10}\times 10^4$. As $N$ is increased, the time interval between $t_c^+(N)-t_c^-(N)$ becomes smaller and is expected to be zero in the limit $N\to \infty$. 
The model parameter values are $\lambda=2.4$ and $g=1/2$. (b) Plot of the order parameter $m(t;N)$ as a function of time $t=L/N$ for different $g$ values but a fixed $\lambda=2.4$. As $g$ is increased, the suppression strength  becomes weaker and thus the transition point $t_c$ occurs earlier. When $g=1$, the extreme case, the model reduces to the original static model, and the transition point occurs at $t=0$.
}
\label{fig:orderparameter}
\end{figure}

Here we investigate properties of a PT in the $r$-SF model. We take the order parameter as the fraction of nodes belonging to a giant cluster, which is denoted as $m(t)$, where $t\equiv L/N$. For finite systems, the order parameter increases very slowly in the early time regime, very rapidly in the intermediate time regime, and gradually in the late time regime, as shown in Fig.~\ref{fig:orderparameter}(a) and (b). 


\begin{figure}[!btp]
\centering
\includegraphics[width=1.0\linewidth]{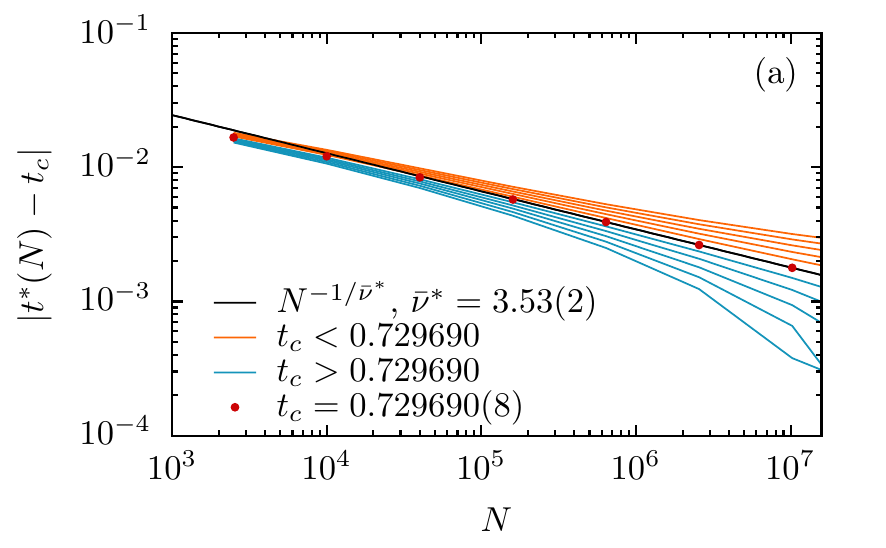}
\includegraphics[width=1.0\linewidth]{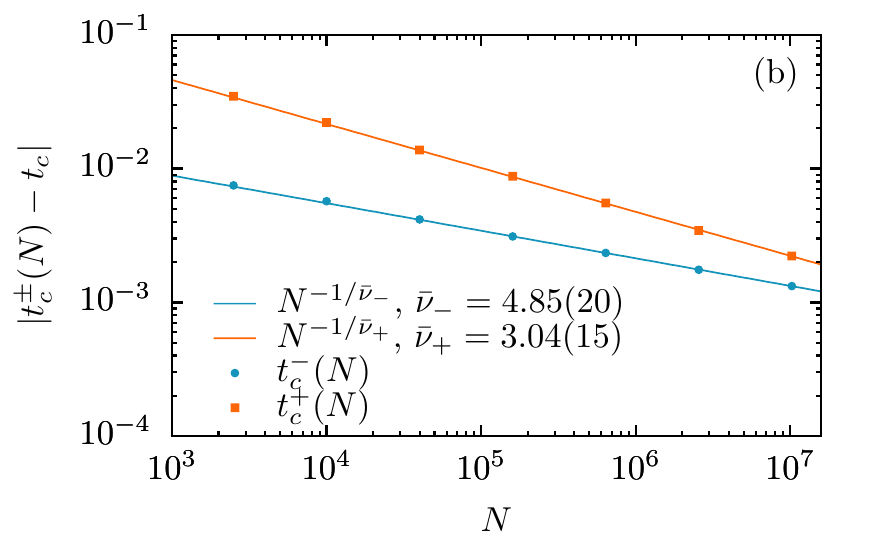}
\caption{
(a) Determination of the transition point $t_c$ in the thermodynamic limit. Using $t^*(N)$ values measured from Fig.~\ref{fig:orderparameter}, we determine $t_c$ as the value satisfying the power-law behavior $|t^*(N)-t_c|\sim N^{-1/\bar \nu^*}$. (b) Determination of the associated critical exponents $\bar \nu_{\pm}$ using the simulation data at the upper and lower percolation thresholds $t_{c}^{\pm}(N)$. In the simulations, $\lambda=2.4$, and $g=1/2$ are used.
}
\label{fig:threshold}
\end{figure}

To determine a transition point $t_c$ in the limit $N\to \infty$, for a fixed system size $N$ and each realization, we obtain the order parameter $m_e(t;N)$ as a function of $t$, where the subscript $e$ represents a single realization. We determine the $t_e^*(N)$ as the point at which the slope $dm_e/dt$ becomes maximum. Next, we take the ensemble averages of $m_e$ and $t_e^*$, and obtain $m(t;N)$ and $t^*(N)$. In the thermodynamic limit, $t^*(N)$ is expected to converge to the transition point $t_c$. Thus, assuming that the finite-size scaling approach works, we determine $t_c$ as the value satisfying the power-law relation $\left|t_c - t^{*}(N)\right| \sim N^{-1/\bar\nu^*}$, where the exponent $\bar \nu^*$ is also obtained [Fig.~\ref{fig:threshold}(a)]. 

\begin{figure}[!btp]
	\centering
	\includegraphics[width=1.0\linewidth]{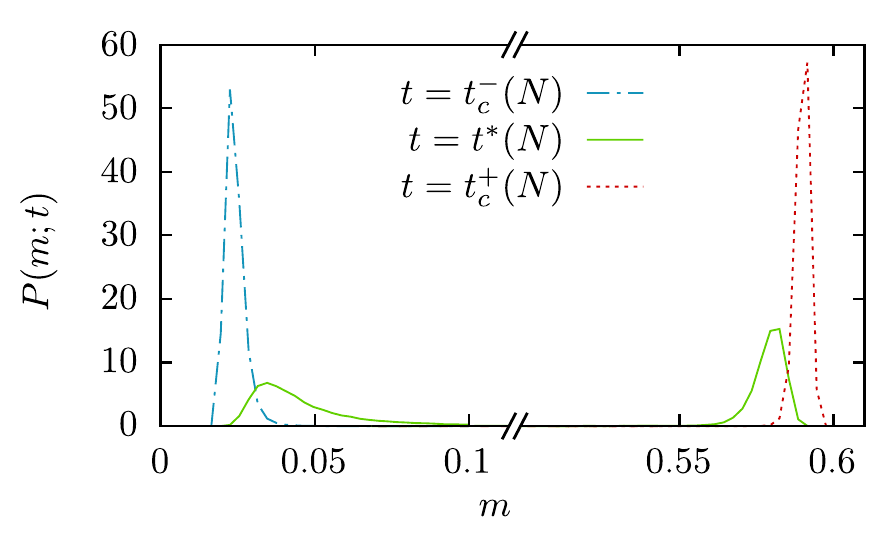}
	\caption{Evolution of the distribution of the order parameter for $g=1/2$. The data are obtained for $\lambda=2.4$ and $N=2^{10} \times 10^{4}$ at  
		$t_{c}^{-}(N)=0.731023$, $t^*(N)=0.731468$, and $t_c^+(N)=0.731924$.}
	\label{fig:order_distribution}
\end{figure}

Next, to check whether the PT is continuous or discontinuous, we first measure the distribution of the order parameter $m$, denoted as \Pm,, by accumulating them over different configurations as a function of $m$ for fixed values of $t$ and $N$. Then we find that there exist two time steps, $t_c^-(N)$ and $t_c^+(N)$. For $t \le t_c^-(N)$, \Pm, exhibits a peak at a certain $m$ near $m=0$, which is denoted as $m^-(N)$. In addition, $m_*^-(N)$ is the $m^-(N)$ value at a particular time $t_c^-(N)$, at which \Pm, begins to exhibit another peak at $m^+(N)$ near $m = 1$, as shown in Fig.~\ref{fig:order_distribution}. As $t$ is increased further, the peak at $m^-$ becomes smaller, whereas the other peak at $m^+$ becomes larger. At $t=t_c^+(N)$, the peak at $m^-$ disappears, and only the peak at $m_*^+$ remains. For $t > t_c^+(N)$, the peak at $m^+(N)$ grows with $t$. As $N$ is increased, $t_c^-(N)$ and $t_c^+(L)$ converge to a certain value $t_c$ [Fig.~\ref{fig:threshold}(b)]; in addition, $m^-(N)\to 0$, and $m^+(N)$ approaches a certain value $m_0$. This result suggests that the order parameter exhibits a jump of size $m_0$ at $t_c$ in the thermodynamic limit. Moreover, we find a finite-size scaling behavior that
\begin{align}\label{scaling_tc}
|t_c^\pm(N)-t_c|\sim N^{-1/\bar \nu_\pm}. 
\end{align} 

\begin{figure}[!tbp]
\centering
\includegraphics[width=1.0\linewidth]{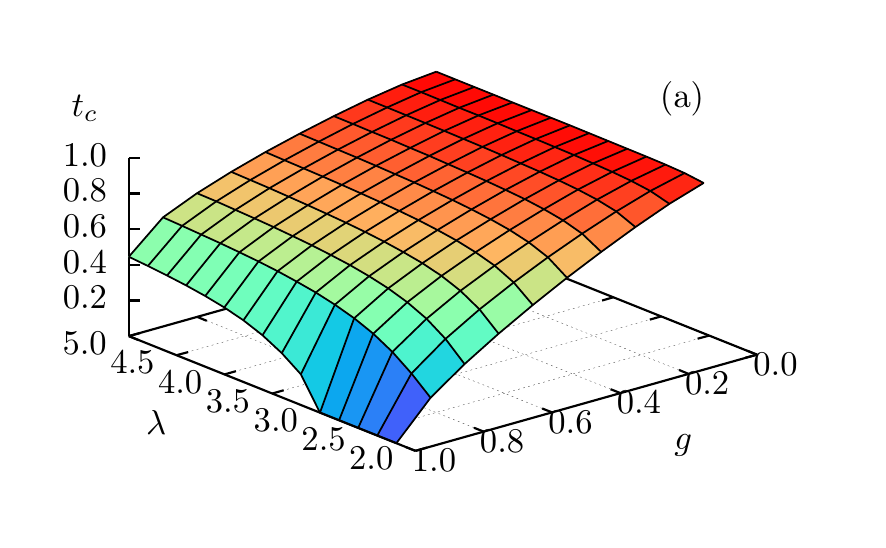}
\includegraphics[width=1.0\linewidth]{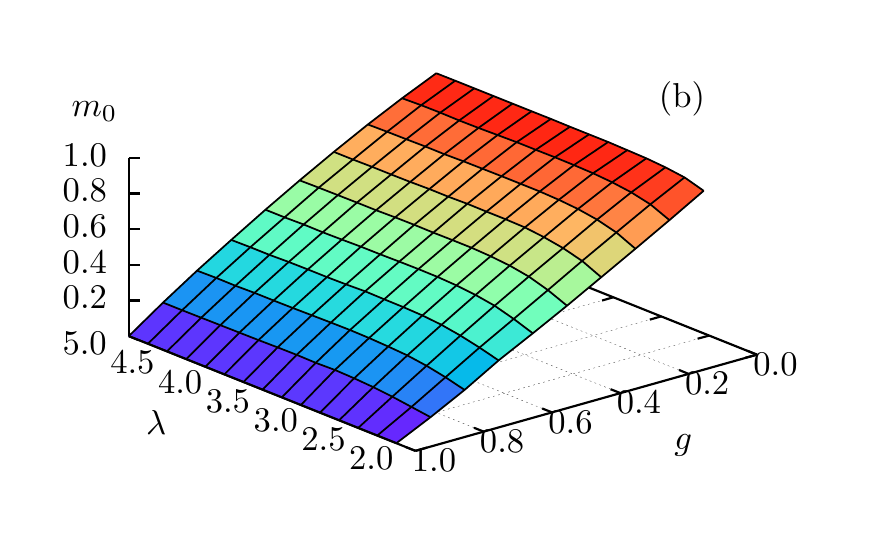}
\caption{
Plots of (a) percolation threshold $t_{c}$ and (b) jump size $m_0$ of the order parameter as a function of degree exponent $\lambda$ and model parameter $g$.}
\label{fig:heatmap}
\end{figure}

The percolation threshold $t_c$ and jump size $m_0$ are depicted as a function of degree exponent $\lambda$ and a model parameter $g$ in Fig.~\ref{fig:heatmap}. Remarkably, when $2 < \lambda < 3$, the transition point $t_c$ is not zero but finite. This behavior originates from the global suppression effect, which usually delays the transition point $t_c$. Here we find that this delay appears even in the SF network with $2 < \lambda < 3$.\\

\begin{figure}[!tbp]
\centering
\includegraphics[width=1.0\linewidth]{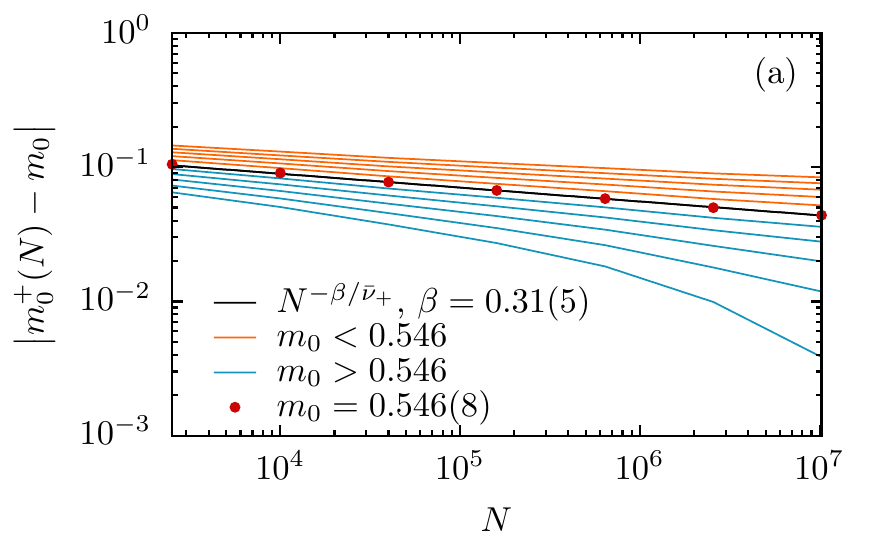}
\includegraphics[width=1.0\linewidth]{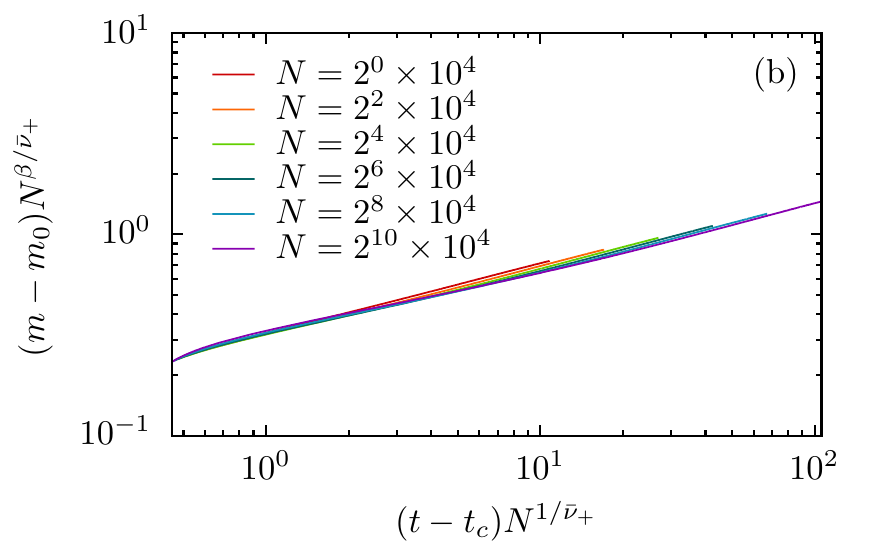}
\caption{
(a) Plot of $\left|m_{0}^{+}(N) - m_{0}\right|$ versus the system size $N$.
(b) Plot of the scaled order parameter $(m-m_{0}) N^{\beta/\bar{\nu}_{+}}$ versus scaled control parameter $(t-t_c)N^{1/\bar{\nu}_{+}}$ for different system sizes. Here, $\lambda=2.4$ and $g=1/2$.
}
\label{fig:order_jump}
\end{figure} 

\section{Critical behavior of the $r$-SF model}\label{sec:CH}
\subsection{The order parameter}

To determine the critical exponent $\bar \nu$ among the exponents $\bar \nu^*$, $\bar \nu_+$, and $\bar \nu_-$ that are associated with the correlation size,   we consider the finite-size scaling behavior of the order parameter. When the system size is increased, the order parameter at the upper percolation threshold, $t_c^+(N)$, behaves as
\begin{align}\label{scaling_m}
	\left| m_{0}^{+}(N) - m_{0}\right| \sim N^{-\beta/\bar{\nu}_{+}}.
\end{align}
In Fig.~\ref{fig:order_jump}(a), we find that this finite-size scaling behavior appears. We also find that the above formula is valid for general $m > m_0$, as shown in Fig.~\ref{fig:order_jump}(b). Thus, $\bar \nu$ is obtained as $\bar \nu_+$. Using this exponent, we determine the critical exponent $\beta$ associated with the order parameter. 


Based on the scaling behaviors Eqs.~\eqref{scaling_tc} and \eqref{scaling_m} for $g < 1$, the order parameter is expressed as follows: 
\begin{align}\label{eq:op}
	m(t) = \begin{cases}
		0 & (t < t_{c}),\\
		m_{0} + a(t-t_{c})^{\beta} & (t \geq t_{c}),
	\end{cases}
\end{align}
where $a$ and $m_{0}$ are constants. Therefore, the PT in the $r$-SF model is hybrid. 

\begin{figure}[!h]
\centering
\includegraphics[width=1.0\linewidth]{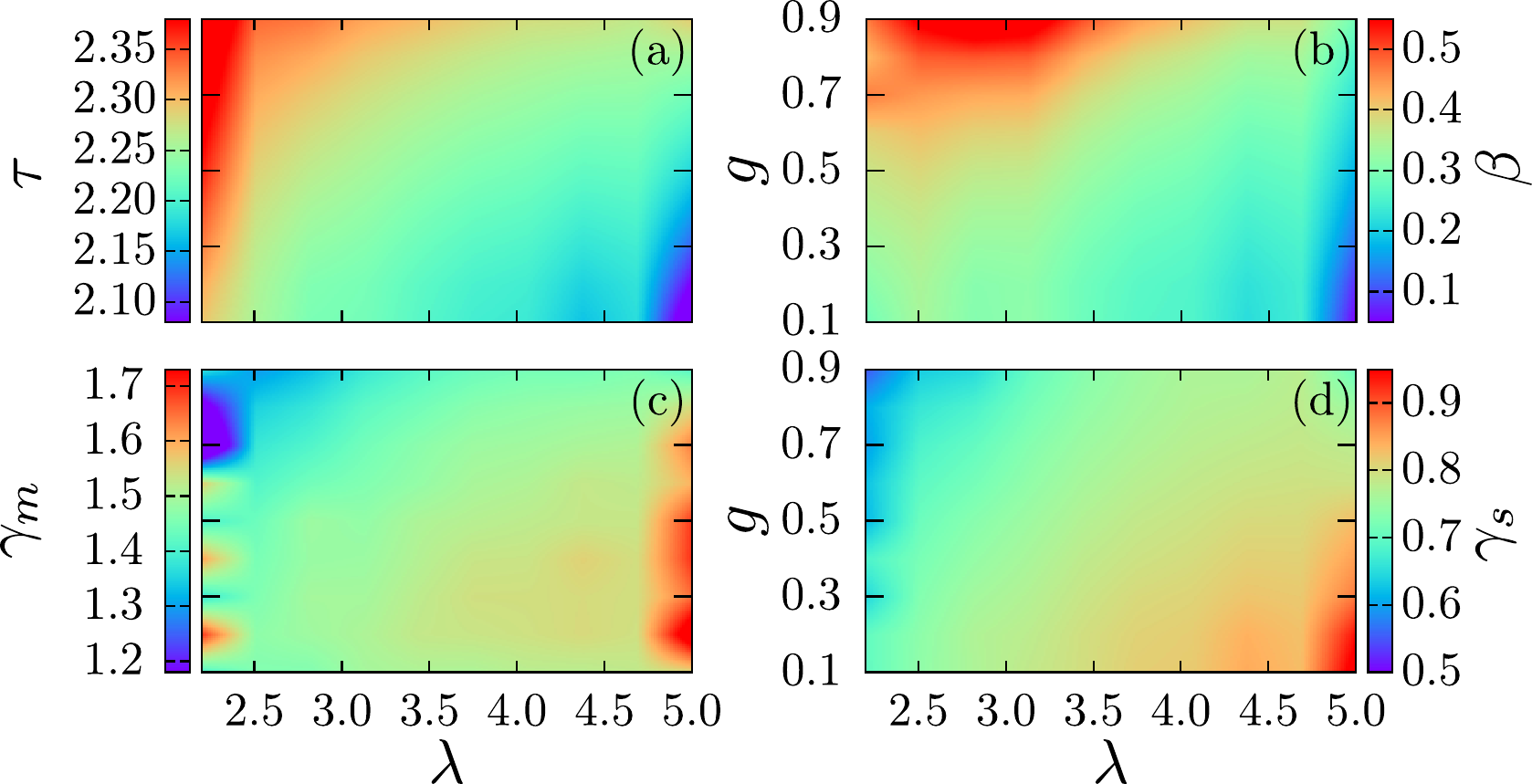}
\caption{
Graphical representation of TABLE~\ref{tab:data}.
The critical exponents (a) $\tau$, (b) $\beta$, (c) $\gamma_{m}$, and (d) $\gamma_{s}$ vary depending on the degree exponent $\lambda$ and global suppression strength $g$.
\label{fig:data}
}
\end{figure}

In contrast to the case of conventional SF networks, analytic solutions of the critical exponents are not feasible for this $r$-SF model. Instead, we perform extensive numerical simulations and determine the critical exponents $\beta$ and $\bar \nu$ for various values of $g$ and $\lambda$ by applying the finite-size scaling analysis introduced in Sec.~III. The numerical values of $\beta$ and $\bar \nu$ for various values of $g$ and $\lambda$ are listed in Table I and depicted in Fig.~\ref{fig:data}.

In addition, we measure the exponent $\chi_m$ associated with the susceptibility induced by sample-to-sample fluctuations of the order parameter; it is defined as 
\begin{align}
	\chi_{m}=N \left[\langle m_e^{2} \rangle - m^2 \right] \sim (t-t_c)^{-\gamma_m} 
	\label{eq:susc}
\end{align}
for $t > t_c$. The exponent $\gamma_m$ values for various $\lambda$ and $g$ are listed in Table I and depicted in Fig.~\ref{fig:data}.

\begin{table*}[h]
	\centering
	\bgroup
	\def\arraystretch{1.4}
	\begin{tabular}{ c c c c c c c c c c c c c}
		\hline
		\hline
		$\lambda$ & $g$ & $t_c$ & $m_{0}$ & $\bar{\nu}_{+}$ & $\tau$ & $\beta$ & $\gamma_{m}$ & $\gamma_{s}$  \\ 
		\hline
		$2.2$ & $0.3$ & $0.83406(3)$ & $0.713(8)$ & $4.96(20)$ & $2.331(8)$ & $0.33(5)$ & $1.38(4)$ & $0.64(40)$ \\
		$2.2$ & $0.5$ & $0.65725(3)$ & $0.501(20)$ & $4.88(30)$ & $2.371(8)$ & $0.37(20)$ & $1.40(14)$ & $0.62(40)$ \\
		$2.2$ & $0.7$ & $0.4522(1)$ & $0.302(18)$ & $4.84(30)$ & $2.408(8)$ & $0.47(30)$ & $0.97(22)$ & $0.60(20)$ \\
		$2.4$ & $0.3$ & $0.879164(5)$ & $0.762(5)$ & $3.02(20)$ & $2.266(5)$ & $0.27(5)$ & $1.49(3)$ & $0.73(30)$ \\
		$2.4$ & $0.5$ & $0.729690(8)$ & $0.546(8)$ & $3.04(15)$ & $2.315(3)$ & $0.31(5)$ & $1.59(23)$ & $0.68(15)$ \\
		$2.4$ & $0.7$ & $0.54559(2)$ & $0.331(13)$ & $3.16(10)$ & $2.361(3)$ & $0.36(10)$ & $1.32(5)$ & $0.64(20)$ \\
		$2.6$ & $0.3$ & $0.902145(1)$ & $0.791(8)$ & $2.39(15)$ & $2.229(5)$ & $0.23(10)$ & $1.43(11)$ & $0.74(20)$ \\
		$2.6$ & $0.5$ & $0.771438(8)$ & $0.579(10)$ & $2.40(15)$ & $2.277(3)$ & $0.28(8)$ & $1.54(10)$ & $0.73(10)$ \\
		$2.6$ & $0.7$ & $0.606588(5)$ & $0.356(2)$ & $2.47(3)$ & $2.324(3)$ & $0.33(2)$ & $1.50(15)$ & $0.68(3)$ \\
		$2.8$ & $0.3$ & $0.915666(3)$ & $0.807(3)$ & $2.22(30)$ & $2.202(3)$ & $0.20(5)$ & $1.62(2)$ & $0.80(20)$ \\
		$2.8$ & $0.5$ & $0.7976023(8)$ & $0.597(8)$ & $2.15(13)$ & $2.252(5)$ & $0.26(5)$ & $1.68(19)$ & $0.76(5)$ \\
		$2.8$ & $0.7$ & $0.647359(5)$ & $0.373(13)$ & $2.29(20)$ & $2.298(5)$ & $0.29(8)$ & $1.41(1)$ & $0.70(15)$ \\
		$3.0$ & $0.3$ & $0.9244142(3)$ & $0.816(3)$ & $2.16(40)$ & $2.185(5)$ & $0.19(3)$ & $1.57(6)$ & $0.81(18)$ \\
		$3.0$ & $0.5$ & $0.8151672(5)$ & $0.607(10)$ & $2.12(18)$ & $2.233(8)$ & $0.23(5)$ & $1.54(1)$ & $0.76(10)$ \\
		$3.0$ & $0.7$ & $0.6756370(8)$ & $0.383(10)$ & $2.23(13)$ & $2.280(5)$ & $0.28(5)$ & $1.47(3)$ & $0.71(20)$ \\
		$3.2$ & $0.3$ & $0.9304640(8)$ & $0.820(2)$ & $2.18(5)$ & $2.171(8)$ & $0.17(1)$ & $1.56(10)$ & $0.82(18)$ \\
		$3.2$ & $0.5$ & $0.8275858(3)$ & $0.612(5)$ & $2.11(30)$ & $2.220(8)$ & $0.22(3)$ & $1.56(1)$ & $0.79(20)$ \\
		$3.2$ & $0.7$ & $0.695979(2)$ & $0.387(8)$ & $2.15(18)$ & $2.266(8)$ & $0.27(3)$ & $1.54(7)$ & $0.75(5)$ \\
		$3.4$ & $0.3$ & $0.9348630(8)$ & $0.820(10)$ & $2.08(30)$ & $2.159(10)$ & $0.16(5)$ & $1.63(6)$ & $0.84(10)$ \\
		$3.4$ & $0.5$ & $0.8367347(8)$ & $0.612(8)$ & $2.10(8)$ & $2.207(8)$ & $0.21(3)$ & $1.66(8)$ & $0.80(5)$ \\
		$3.4$ & $0.7$ & $0.7111001(2)$ & $0.388(5)$ & $2.15(3)$ & $2.255(10)$ & $0.26(3)$ & $1.55(7)$ & $0.75(5)$ \\
		$3.6$ & $0.3$ & $0.9381853(2)$ & $0.820(5)$ & $2.10(40)$ & $2.148(10)$ & $0.15(3)$ & $1.71(1)$ & $0.88(20)$ \\
		$3.6$ & $0.5$ & $0.8436987(5)$ & $0.611(8)$ & $2.07(30)$ & $2.198(10)$ & $0.20(3)$ & $1.65(5)$ & $0.81(20)$ \\
		$3.6$ & $0.7$ & $0.7226563(5)$ & $0.389(8)$ & $2.17(20)$ & $2.247(10)$ & $0.25(3)$ & $1.50(1)$ & $0.76(5)$ \\
		$3.8$ & $0.3$ & $0.9407660(2)$ & $0.819(5)$ & $2.11(40)$ & $2.138(10)$ & $0.14(2)$ & $1.72(1)$ & $0.87(30)$ \\
	    $3.8$ & $0.5$ & $0.849149(1)$ & $0.612(5)$ & $2.14(15)$ & $2.191(10)$ & $0.19(2)$ & $1.56(6)$ & $0.82(8)$ \\
		$3.8$ & $0.7$ & $0.731684(3)$ & $0.386(5)$ & $2.15(18)$ & $2.240(10)$ & $0.24(2)$ & $1.58(6)$ & $0.75(5)$ \\
		$4.0$ & $0.3$ & $0.9428249(2)$ & $0.818(8)$ & $2.06(13)$ & $2.132(10)$ & $0.14(2)$ & $1.68(6)$ & $0.86(20)$ \\
		$4.0$ & $0.5$ & $0.8534892(3)$ & $0.611(5)$ & $2.17(18)$ & $2.187(13)$ & $0.19(2)$ & $1.61(2)$ & $0.82(18)$ \\
		$4.0$ & $0.7$ & $0.7389024(5)$ & $0.384(3)$ & $2.12(20)$ & $2.234(13)$ & $0.23(1)$ & $1.56(3)$ & $0.77(13)$ \\
		$4.2$ & $0.3$ & $0.9444929(3)$ & $0.815(5)$ & $2.10(3)$ & $2.130(13)$ & $0.13(2)$ & $1.78(4)$ & $0.88(10)$ \\
		$4.2$ & $0.5$ & $0.8570188(3)$ & $0.608(8)$ & $2.12(13)$ & $2.183(13)$ & $0.18(2)$ & $1.59(4)$ & $0.81(20)$ \\
		$4.2$ & $0.7$ & $0.7447390(5)$ & $0.383(5)$ & $2.17(30)$ & $2.230(13)$ & $0.23(2)$ & $1.49(6)$ & $0.77(13)$ \\
		$4.4$ & $0.3$ & $0.9458725(1)$ & $0.815(8)$ & $2.15(18)$ & $2.124(13)$ & $0.12(2)$ & $1.60(16)$ & $0.86(13)$ \\
		$4.4$ & $0.5$ & $0.8599336(3)$ & $0.608(5)$ & $2.13(15)$ & $2.181(13)$ & $0.18(2)$ & $1.56(9)$ & $0.83(10)$ \\
		$4.4$ & $0.7$ & $0.7495483(2)$ & $0.382(5)$ & $2.17(30)$ & $2.228(13)$ & $0.23(2)$ & $1.53(2)$ & $0.77(10)$ \\
		$4.6$ & $0.3$ & $0.9470265(5)$ & $0.813(10)$ & $2.16(18)$ & $2.122(13)$ & $0.12(3)$ & $1.62(14)$ & $0.87(5)$ \\
		$4.6$ & $0.5$ & $0.8623685(3)$ & $0.607(10)$ & $2.19(40)$ & $2.178(15)$ & $0.18(3)$ & $1.58(7)$ & $0.83(20)$ \\
		$4.6$ & $0.7$ & $0.7535534(3)$ & $0.380(8)$ & $2.10(2)$ & $2.226(15)$ & $0.23(2)$ & $1.62(7)$ & $0.78(5)$ \\
		$4.8$ & $0.3$ & $0.9480027(2)$ & $0.812(8)$ & $2.11(40)$ & $2.121(15)$ & $0.12(3)$ & $1.62(14)$ & $0.88(20)$ \\
		$4.8$ & $0.5$ & $0.8644256(2)$ & $0.606(10)$ & $2.19(30)$ & $2.176(13)$ & $0.18(3)$ & $1.59(6)$ & $0.84(15)$ \\
		$4.8$ & $0.7$ & $0.7569228(3)$ & $0.380(8)$ & $2.12(30)$ & $2.226(15)$ & $0.22(2)$ & $1.60(5)$ & $0.78(5)$ \\
		$r$-{ER} & $0.3$ & $0.9570(2)$ & $0.81(5)$  &            & $2.12(4)$   & $0.13(5)$ &           & $0.88(5)$ \\ 
		$r$-{ER} & $0.5$ & $0.8826(2)$ & $0.62(5)$  &            & $2.18(4)$   & $0.21(5)$ &           & $0.83(5)$ \\ 
		$r$-{ER} & $0.7$ & $0.7852(3)$ & $0.41(5)$  &            & $2.22(4)$   & $0.28(5)$ &           & $0.81(5)$ \\ 
		\hline
		\hline
	\end{tabular}
\caption{
Percolation threshold $t_{c}$, initial fraction of the giant cluster $m_{0}$, and critical exponents of the $r$-SF networks for various degree exponents $\lambda$ and model parameters $g$. The numerical values for $r$-ER are adopted from Ref.~\cite{Cho2016}. 
\label{tab:data}
}
\egroup
\end{table*}

\subsection{Size distribution of finite clusters}

\begin{figure}[!tbp]
\centering
\includegraphics[width=1.0\linewidth]{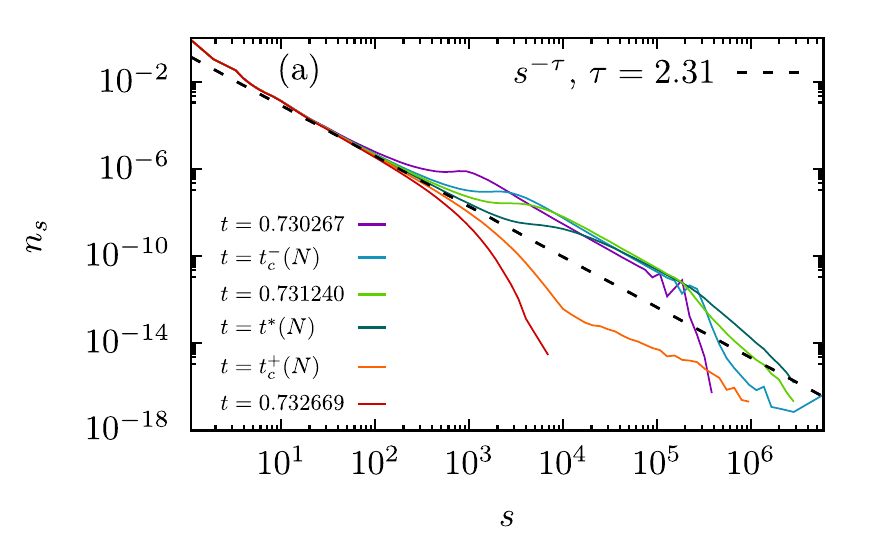}
\includegraphics[width=1.0\linewidth]{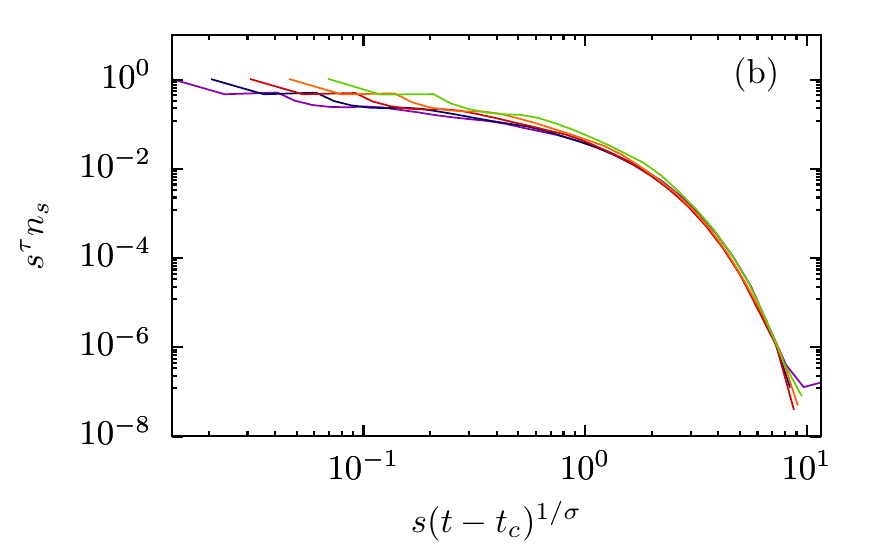}
\caption{
(a) Plot of the cluster size distribution $n_{s}$ versus cluster size $s$ for finite clusters. It exhibits power-law decay for small $s$ but a bump in the tail. The power-law regime is extended as $t$ approaches $t_c$. For $t \ge t_c^+(N)$, the bump disappears and the $n_s(t)$ exhibits exponential decay beyond the power-law regime.  
(b) Scaling behavior of the cluster size distribution with the critical exponent $\sigma =1$. The numerical data are obtained using the parameter values $\lambda=2.4$ and $g=0.5$. The system size is taken as $N=2^{10}\times 10^{4}$. Recall that  $t_{c}^{-}(N)=0.731023$, $t^*(N)=0.731468$, and $t_c^+(N)=0.731924$.
}
\label{fig:ns}
\end{figure}

We consider the cluster-size distribution $n_{s}$, that is, the number of clusters of size $s$ divided by the system size $N$. For $t < t_c^-(N)$, $n_s$ exhibits power-law decay as $n_s\sim s^{-\tau}$ up to a characteristic size $s^*$, beyond which it exhibits exponential decay. Near $t_c^-(N)$, a bump suddenly appears in the tail, and the position of the bump is characterized by $s^*$. During the interval $t_c^-(N) < t < t_c^+(N)$, as time passes, the bump moves toward larger cluster size and becomes wider and smaller. Ultimately, at $t_c^+(N)$, the bump disappears completely; the distribution of finite clusters exhibit power-law decay as $n_{s}\sim s^{-\tau}$, and the giant cluster of size $m_0^+(N)N$ is generated. After $t_c^+(N)$, $n_s$ of finite clusters exhibits typical supercritical behavior; i.e., it exhibits power-law decay up to a characteristic size $s_c$, after which the tail decays exponentially. This evolution process is shown in Fig.~\ref{fig:ns}(a). In addition, $s_c$ is scaled as $s_c\sim (t-t_{c})^{-1/\sigma}$ as shown in Fig.~\ref{fig:ns}(b). The exponent $\sigma$ is measured as $\sigma \approx 1$ regardless of $\alpha$ and $g$. This behavior is universal for the HPT owing to the jump in the order parameter~\cite{DaCosta2014,DaCosta2015}. 
We checked that the scaling relation $\beta=(\tau-2)/\sigma$ holds in the supercritical regime. However, $\bar{\nu}^{*}$ and $\bar{\nu}_{\pm}$ do not satisfy the hyperscaling relation $\bar{\nu}_{\pm} \neq (\tau - 1)/\sigma$.

\begin{figure}[!tbp]
	\centering
	\includegraphics[width=1.0\linewidth]{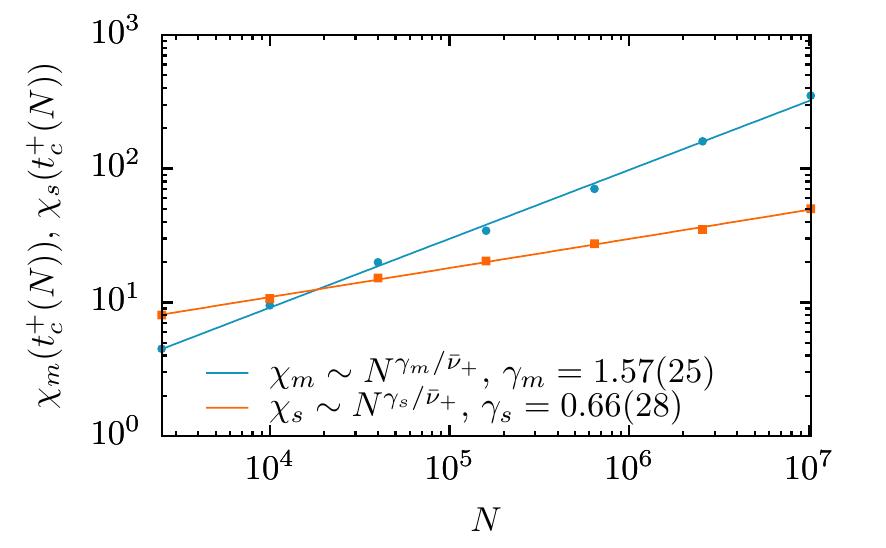}
	\caption{
		Fluctuation of the order parameter $\chi_{m}$ and the average cluster size $\chi_{s}$ at the upper percolation threshold. Here, $\lambda=2.4$ and $g=1/2$.
	}
	\label{fig:chi}
\end{figure}

We consider the mean cluster size distribution, which is defined as 
\begin{align}
	\chi_{s} \equiv \sum s^{2}n_{s} / \sum s n_{s}.
\end{align}
It diverges as 
\begin{align}
\chi_{s} \sim \left(t-t_{c}\right)^{-\gamma_{s}}.
\end{align}
This behavior is similar to that of the susceptibility defined in Eq.~\eqref{eq:susc} for continuous PTs such as the original PT in random SF networks. However, the exponent value of $\gamma_s$ differs from $\gamma_m$ in this HPT.  The exponent $\gamma_s$ values for various $\lambda$ and $g$ are listed in Table I and depicted in Fig.~\ref{fig:data}. 

Using the finite-size scaling of the upper percolation threshold in Fig.~\ref{fig:chi}, we measure the critical exponent $\gamma_{s}$ for different $\mu$ and $g$. We find that the values of $\gamma_{m}$ and $\gamma_{s}$ depend on $\mu$ and $g$. The exponent $\gamma_{s}$ satisfies the scaling relation $\gamma_{s} = (3-\tau)/\sigma$ for finite clusters. The different behaviors of the two susceptibilities also appear in the restricted percolation model in two dimensions~\cite{Choi2017} and in random newtorks~\cite{Cho2016}. This also seems to be valid for $r$-SF model.

\section{Summary and Discussion}\label{sec:summary}
The restricted static model, which is equivalent to the $r$-SF model, is a generalization of the static model obtained by assigning a global suppression rule to the original static model. We find that even though the global suppression is present, the degree distribution of the $r$-SF model does not change. However, the PT type changes from second-order to hybrid. Thus, the order parameter jumps at a transition point and exhibits critical behaviors for the order parameter and the susceptibility. Moreover, the transition point is delayed. Accordingly, an HPT occurs at a finite transition point $t_c$ even for the degree exponent $2< \lambda < 3$. Also, we can control $t_c$ by the suppression strength $g$. As the global suppression strength $1-g$ is increased, the transition point and the jump size are increased. We determined the critical exponents of the PT by performing extensive simulations in the parameter space ($\lambda$, $g$). These results differ from the previous result obtained from the explosive percolation model~\cite{yscho2009,santo}: the degree exponent is changed by suppression effect, and the transition point is zero for $2 < \lambda < \lambda_c$, where $2.3 < \lambda_c < 2.4$. Our numerical results are listed in Table I and are depicted in Fig.~\ref{fig:data}.


When $g=1$, the $r$-SF model reduces to the static model of ordinary SF networks. In this case, the transition point and critical behaviors depend on the degree exponent $\lambda$, or $\mu=1/(\lambda-1)$. We classify the critical behaviors into three cases:
(i) $\mu \leq 1/3$ ($4 < \lambda$), (ii) $1/3 <\mu <1/2$ ($3 < \lambda < 4$), and (iii) $1/2 < \mu < 1$ ($2 <  \lambda < 3$). For each case, the transition point $t_c$ and critical exponent values are determined analytically as a function of $\lambda$~\cite{lee04}; the results are listed in Table II. Note that the transition point is $t_c=0$ for the case (III).

\begin{table}[ht]
	\centering
	\bgroup
	\def\arraystretch{1.5}
	\begin{tabular}{ c c c c c c c }
		\hline
		\hline
		& $t_c$ & $\tau$ & $\sigma$ & $\beta$ & $\gamma$ & $\bar{\nu}$  \\ 
		\hline
		$4 < \lambda $ & $\frac{(\lambda-3)(\lambda-1)}{2(\lambda - 2)^{2}}$ & $\frac{5}{2}$ & $\frac{1}{2}$ & 1 & 1 & 3 \\
		$3 < \lambda < 4$ & $\frac{(\lambda-3)(\lambda-1)}{2(\lambda - 2)^{2}}$ & $\frac{2\lambda -3}{\lambda-2}$ & $\frac{\lambda-3}{\lambda-2}$ & $\frac{1}{\lambda-3}$ & 1 & $\frac{\lambda-1}{\lambda-3}$ \\
		$2 < \lambda < 3$ & 0 & $\frac{2\lambda -3}{\lambda-2}$ & $\frac{3-\lambda}{\lambda-2}$ & $\frac{1}{3-\lambda}$ & -1 & $\frac{\lambda-1}{3-\lambda}$ \\
		\hline
		\hline
	\end{tabular}
	\caption{
		Critical exponents of the PT on scale-free networks.
		\label{tab:sf}
	}
	\egroup
\end{table} 

For the ordinary SF networks, the case $\lambda > 4$ is regarded as the mean-field limit, in which the critical exponents are independent of $\lambda$. Here, we compare our numerical values of the critical exponents $\tau$, $\beta$, and $\gamma_s$ for $\lambda > 4$ with those for $r$-ER model obtained in Ref.~\cite{Cho2016}. These values are also listed in Table I. Indeed, the exponent values for $\lambda > 4$ seem to be independent of $\lambda$ and close to those of $r$-ER model for respective $g$. Moreover, the exponent of $\tau$ for $r$-SF model with $g < 1$ is smaller than that for the original static SF model, corresponding to the $r$-SF model with $g=1$. This means that the size of the largest cluster for $r$-SF model is larger than that of the original SF model. This is counter-intuitive, because the growth of large cluster is suppressed in $r$-SF model. As the transition point is delayed, more links are added to the system without forming a giant cluster up to $t_c$. This increases the sizes of large clusters. Thus, the size of the largest cluster for $r$-SF model is larger than that of the original SF model in the subcritical regime.  

Recently, the underlying mechanism of the HPT in the $r$-ER model was investigated~\cite{Park2019}. As clusters are merged, the elements of each partition may be updated. In this process, clusters may move back and forth across the set boundary. Thereby, each node may also move similarly. This dynamics may be regarded as a tug-of-war-like process. The duration times that each node stays in one set until moving to the other have the distribution that exhibits power-law behavior. We expect that a similar dynamics occurs in the $r$-SF model. In a fundamental aspect, the asymmetric rules in the restricted ER and SF models generate a symmetry-breaking factor that plays a role of the negative cubic term in the Landau free energy scheme. Moreover, the system maintains criticality through the tug-of-war-like process with a global information. By combining these two factors, a HPT occurs in $r$-SF model.

\begin{acknowledgments}
This work was supported by the NRF of Republic of Korea, Grant No.~NRF-2014R1A3A2069005 (BK). \\

{\bf Data availability}: Not available.
\end{acknowledgments}

\end{document}